\documentclass[manuscript,screen]{acmart}
\AtBeginDocument{%
  }

\setcopyright{acmlicensed}
\copyrightyear{2018}
\acmYear{2018}
\acmDOI{XXXXXXX.XXXXXXX}
\acmConference[Conference acronym 'XX]{Make sure to enter the correct
  conference title from your rights confirmation email}{June 03--05,
  2018}{Woodstock, NY}

\acmISBN{978-1-4503-XXXX-X/18/06}

\usepackage{enumitem}
\usepackage{pifont}
\usepackage{multirow}
\usepackage{tcolorbox}
\usepackage{color,xcolor}
\usepackage{listings,amsfonts}
\usepackage{caption}
\usepackage{subcaption}
\usepackage{threeparttable}
\usepackage{bbding}
\usepackage{pdflscape}
\usepackage{graphicx}
\usepackage{booktabs} 
\usepackage{longtable}
\usepackage{flushend}
\usepackage{cleveref}
\usepackage{fancybox}
\usepackage{rotating}
\usepackage{booktabs}
\usepackage{makecell}
\usepackage{tabularx}  

\begin{document}

\title{Foundation Models for Autonomous Driving Systems: An Initial Roadmap}

\author[X Wu]{Xiongfei Wu}
\affiliation{%
    \institution{University of Luxembourg}
    \country{Luxembourg}}

\author[M Cheng]{Mingfei Cheng}
\affiliation{%
    \institution{Singapore Management University}
    \country{Singapore}}

\author[X Ren]{Xiaoning Ren}
\affiliation{
    \institution{University of Science and Technology of China}
    \country{China}}

\author[Q Hu]{Qiang Hu$^*$}\thanks{$^*$ Corresponding author: Qiang Hu.}
\affiliation{%
    \institution{Tianjin University}
    \country{China}}

\author[J Chen]{Jianlang Chen}
\affiliation{
    \institution{Kyushu University}
    \country{Japan}}

\author[Y Huang]{Yuheng Huang}
\affiliation{
    \institution{The University of Tokyo}
    \country{Japan}}

\author[M Cordy]{Maxime Cordy}
\affiliation{
    \institution{University of Luxembourg}
    \country{Luxembourg}}

\author[Y Zhang]{Yao Zhang}
\affiliation{
    \institution{Tianjin University}
    \country{China}}

\author[X Xie]{Xiaofei Xie}
\affiliation{%
 \institution{Singapore Management University}
  \country{Singapore}}

\author[L Ma]{Lei Ma}
\affiliation{%
  \institution{The University of Tokyo \& University of Alberta}
  \country{Japan \& Canada}}

\author[Y Traon]{Yves Le Traon}
\affiliation{
    \institution{University of Luxembourg}
    \country{Luxembourg}}

\renewcommand{\shortauthors}{Wu et al.}

\begin{abstract}
Recent advances in foundation models (FMs), including large language models (LLMs), vision-language models (VLMs), and world models, have opened new opportunities for autonomous driving systems (ADSs) in perception, reasoning, decision-making, and interaction. However, ADSs are safety-critical cyber-physical systems, and integrating FMs into them raises substantial software engineering challenges in data curation, system design, deployment, evaluation, and assurance. To clarify this rapidly evolving landscape, we present an initial roadmap, grounded in a structured literature review, for integrating FMs into autonomous driving across three dimensions: FM infrastructure, in-vehicle integration, and practical deployment. For each dimension, we summarize the state of the art, identify key challenges, and highlight open research opportunities. Based on this analysis, we outline research directions for building reliable, safe, and trustworthy FM-enabled ADSs.
\end{abstract}




\maketitle

\section{Introduction}
\label{sec:intro}

Autonomous driving systems (ADSs) operate in open, dynamic, and safety-critical environments. Traditional approaches, while effective in controlled environments, often struggle with unseen situations, unexpected obstacles, and dynamic interactions that characterize real-world driving conditions~\cite{wu2024prospective}. A key limitation is their reliance on predetermined rules and supervised learning over finite labeled datasets, which cannot fully capture the diversity and long-tail nature of scenarios.

The emergence of foundation models (FMs) trained on vast and diverse datasets has demonstrated unprecedented capabilities in reasoning and generalization across various domains~\cite{bommasani2022on}. These models have exhibited remarkable abilities in understanding context, reasoning about the context, and generating appropriate responses. Their success in natural language processing~\cite{devlin2019bert,touvron2023llama} and computer vision~\cite{radford2021learning, liu2023visual, gpt4v} tasks suggests promising applications in addressing the fundamental challenges of autonomous driving.

The autonomous driving ecosystem stands to benefit significantly from the integration of FMs, which can enhance real-world scenario understanding, improve decision-making, and facilitate robust system development. For instance, FMs can leverage their broad knowledge base to interpret complex traffic scenarios, understand natural language instructions from passengers, and make informed decisions in previously unseen situations~\cite{zhou2024vision}. This integration could bridge the gap between traditional autonomous driving systems' capabilities and the requirements for truly robust autonomous operation in diverse real-world conditions. However, despite their promising enhancement for autonomous driving, FMs also have certain problems. Due to their increasing complexity, FMs heavily rely on \textit{data} and are especially hard to \textit{manage}. This becomes a significant obstacle to integrating them into autonomous driving systems.

\begin{figure*}[t]
    \centering
    \includegraphics[width=\textwidth]{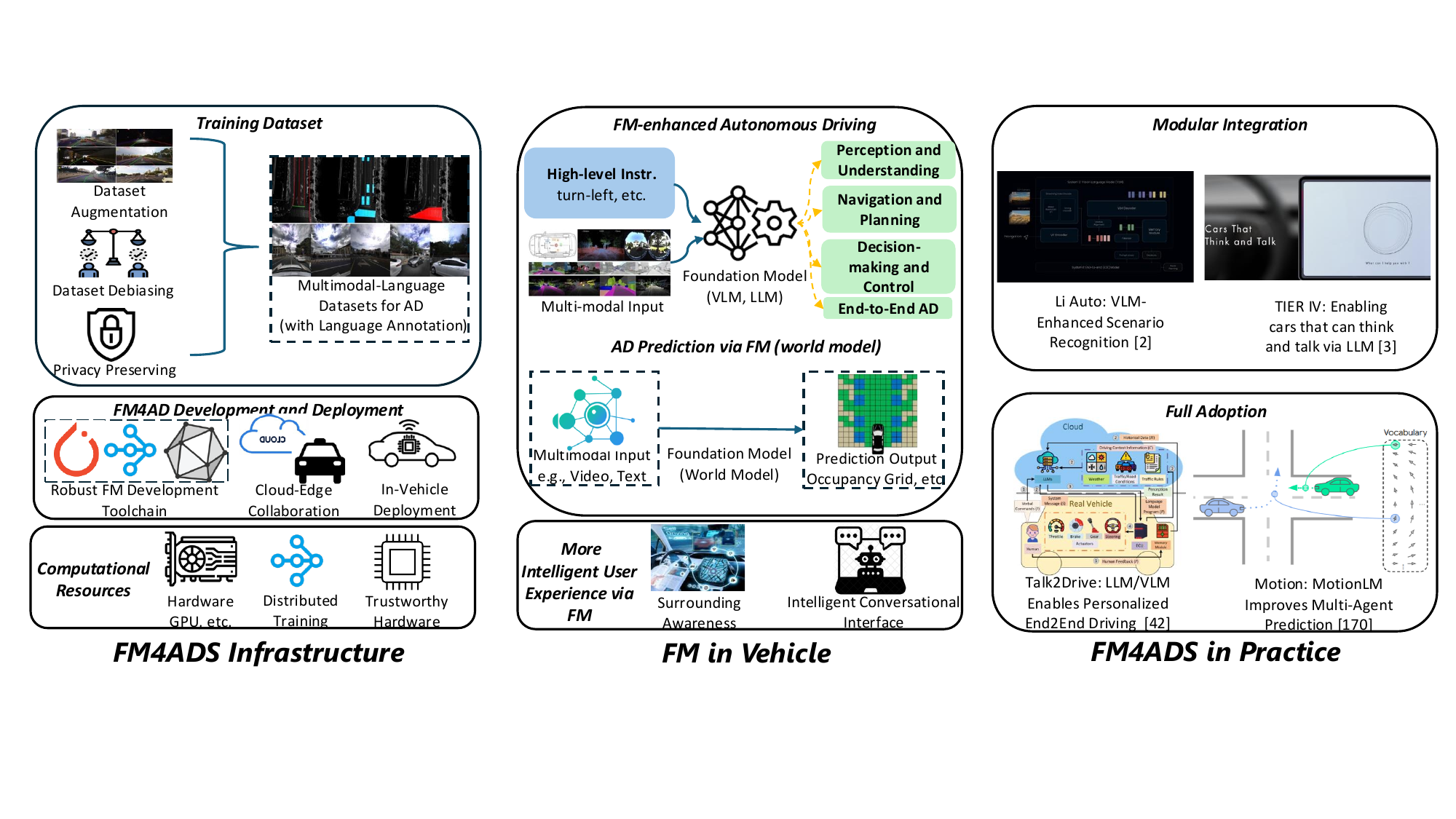}
    \caption{Overview of the roadmap.}
    \Description{Topics covered in this paper.}
    \label{fig:overview}
\end{figure*}

To help researchers better understand the role of foundation models in autonomous driving, considerable efforts have been made to broaden the community’s perspective on their potential contributions. Song~\textit{et al.}~\cite{song2025generative} and Petrovic~\textit{et al.}~\cite{petrovic2024llm} surveyed the use of generative AI, particularly LLMs, for testing autonomous driving systems. Other studies have explored how foundation models can advance autonomous driving more broadly~\cite{wu2024prospective, yan2024forging, gao2025survey}. While these works provide valuable insights and highlight important challenges and opportunities, most focus on specific topics or model families, for example, using LLMs to generate testing scenarios for ADS. To move beyond such fragmented perspectives, we propose a structured roadmap for integrating FMs into autonomous driving. Our roadmap spans three dimensions: FM infrastructure, its integration across autonomous driving system modules, and their practical real-world applications, as shown in Figure~\ref{fig:overview}. For each dimension, we review the current research progress, identify existing challenges, and highlight research gaps that need to be addressed by the community. Beyond a modular overview, Table~\ref{tab:cause_effect_chain} connects the three dimensions by showing how infrastructure-level decisions propagate through in-vehicle risks and ultimately manifest as failures in practice. Through this analysis, we aim to guide future research on building reliable, safe, and trustworthy FM-enabled ADSs.

The main contributions of this paper are summarized as follows:
\begin{itemize}[leftmargin=15pt]
    \item We provide a structured review of integrating FMs into autonomous driving, structured along three critical dimensions: the underlying infrastructure, the in-vehicle system, and practical real-world applications.

    \item For each dimension, we conduct an in-depth analysis to identify key challenges and highlight promising research opportunities, to actively pinpoint critical roadblocks and avenues for innovation.

    \item Based on our analysis, we formulate a forward-looking research roadmap with concrete short, mid, and long-term goals, designed to guide researchers and practitioners in prioritizing their efforts to build the next generation of safe, reliable, and trustworthy autonomous systems.
\end{itemize}

\begin{table*}[t]
\centering
\small
\caption{End-to-End Dependency Analysis: Tracing Infrastructure Decisions to Real-World Practice Failures.}
\label{tab:cause_effect_chain}
\renewcommand{\arraystretch}{1.0} 
\begin{tabularx}{\textwidth}{
    >{\raggedright\arraybackslash\hsize=1.0\hsize}X 
    >{\raggedright\arraybackslash\hsize=1.0\hsize}X 
    >{\raggedright\arraybackslash\hsize=1.0\hsize}X 
    >{\centering\arraybackslash\hsize=0.2\hsize}X } 
\toprule
\textbf{Infrastructure Decision} \par \textit{(Root Cause)} & 
\textbf{Vehicle Risk} \par \textit{(System Effect)} & 
\textbf{Practice Failure} \par \textit{(Final Consequence)} & 
\textbf{Ref.} \\
\midrule

\textbf{Biased Dataset Selection}: \newline Limited diversity in demographic attributes (e.g., age, sex, skin tone). & 
\textbf{Perception Blind Spots}: \newline Inaccurate detection or classification of underrepresented groups; higher "Miss Rates" (MR). & 
\textbf{Systemic Safety Risks}: \newline Higher accident rates for specific populations; ethical violations in real-world deployment. & 
Sec.~\ref{sec:dataset} \newline (Chal.~I) \\
\midrule

\textbf{Unscrubbed PII in Training}: \newline Inclusion of sensitive data (faces, license plates) and lack of "Machine Unlearning" protocols. & 
\textbf{Data Memorization}: \newline The FM inadvertently encodes and retains specific personal information within its parameters. & 
\textbf{Privacy Leaks}: \newline Model regenerates private data during inference, causing GDPR violations or identity leaks. & 
Sec.~\ref{sec:dataset} \newline (Chal.~I) \\
\midrule

\textbf{Data Collection Gaps}: \newline Absence of safety-critical edge cases (e.g., accident aftermath, extreme weather) in training data. & 
\textbf{Reasoning Defects}: \newline Model inability to interpret or react appropriately to rare, high-risk scenarios (Long-tail distribution). & 
\textbf{Safety Validation Failure}: \newline Catastrophic failure when the vehicle encounters unseen dangerous situations. & 
Sec.~\ref{sec:dataset} \newline (Chal.~II) \\
\midrule

\textbf{Lack of Hardware Root-of-Trust}: \newline Deployment on commodity accelerators without TEEs. & 
\textbf{Execution Exposure}: \newline Model weights, intermediate activations, and sensor streams are visible to the host OS. & 
\textbf{Model Theft \& Spoofing}: \newline Theft of proprietary model or injection of fake sensor data into the loop. & 
Sec.~\ref{sec:comutational_resource} \newline (Chal.~I) \\
\midrule

\textbf{Aggressive Model Optimization}: \newline High levels of pruning or quantization to meet on-board efficiency constraints. & 
\textbf{Security Vulnerabilities}: \newline Lowered robustness thresholds; increased susceptibility to tailored hardware-level or bit-flip attacks. & 
\textbf{System Security Breach}: \newline Unauthorized model weight manipulation or parameter leakage leading to loss of control. & 
Sec.~\ref{sec:dev_deploy} \newline (Chal.~II) \\
\midrule

\textbf{Untested FM Integration}: \newline Using FMs without robust multi-modal grounding or verification mechanisms. & 
\textbf{Hallucination}: \newline Generation of "ghost objects," misinterpretation of traffic signs, or non-sensical path planning. & 
\textbf{Erratic Control}: \newline "Phantom braking," sudden swerving, or dangerous maneuvers causing collisions. & 
Sec.~\ref{sec:fm4ad} \newline (Chal.~I) \\
\midrule

\textbf{Weak Safety Alignment}: \newline Insufficient "Red Teaming" or guardrails during the instruction-tuning phase of LLM/VLM. & 
\textbf{Jailbreak Susceptibility}: \newline System processes adversarial prompts (e.g., "Drive aggressively") without rejecting them. & 
\textbf{Reckless Driving}: \newline Vehicle violates traffic rules or safety envelopes to satisfy user commands. & 
Sec.~\ref{sec:intelligent_user} \newline (Chal.~\&~Opp.) \\
\midrule
\textbf{Unverified Code Generation}: \newline Utilizing LLMs-generated code without careful inspection. & 
\textbf{Subtle Logic Defects}: \newline Code is syntactically correct but contains semantic flaws or lacks context-aware safety logic. & 
\textbf{Runtime Software Failure}: \newline Unexpected system crashes or unsafe execution paths during complex interactions. & 
Sec.~\ref{sec:practice} \newline (Chal.~II) \\

\bottomrule
\end{tabularx}
\end{table*}

\begin{figure*}[t]
    \centering
    \includegraphics[width=\textwidth]{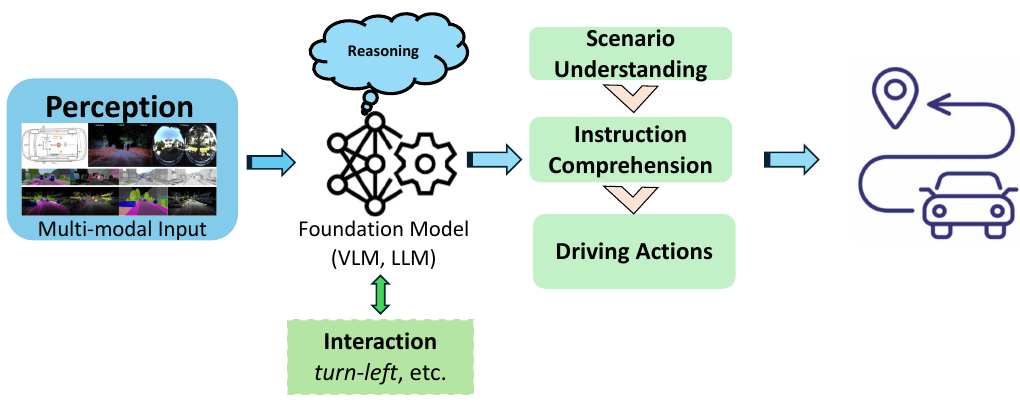}
    \caption{A typical pipeline of enhancing autonomous driving leveraging FMs.}
    \Description{Topics covered in this paper.}
    \label{fig:pipeline}
\end{figure*}

\section{Background, Review Methodology, and Related Work}
\label{sec:priliminary}

In this section, we first provide background on the current state of adopting foundation models in autonomous driving systems to orient the reader. We then describe the systematic review methodology used to select and synthesize the literature. Finally, we position our roadmap with respect to existing surveys and roadmaps, highlighting its scope and distinguishing characteristics.

\subsection{Foundation Model in Autonomous Driving Systems}
Existing integration of FMs into autonomous driving systems can be roughly categorized into \emph{perception and scene understanding}, \emph{navigation and planning}, \emph{decision-making and control}, and \emph{end-to-end autonomous driving}. \Cref{fig:pipeline} shows a typical pipeline of leveraging FMs to enhance autonomous driving. In this section, we briefly discuss representative techniques and identify the challenges and opportunities. For more background and technique details, we refer readers to prior works~\cite{wu2024prospective, gao2025survey, cui2024survey, zhou2024vision}.

\subsubsection{Perception and Scene Understanding}

FMs enhance perception in autonomous driving by enabling context-aware environmental understanding~\cite{zhou2024vision}. VLMs such as LLaVA~\cite{liu2023visual} and GPT-4V~\cite{gpt4v} support tasks like 3D open-vocabulary object detection~\cite{najibi2023unsupervised, pan2024CLIPBEVFormer}, language-guided retrieval~\cite{romero2023zeldavideoanalyticsusing, wei2024BEVCLIP}, and visual question answering (VQA)~\cite{qian2024nuscenesqa, choudhary2024Talk2BEV, nie2025Reason2Drive, liu2024vcd}. Examples include OpenScene~\cite{peng2023openscene} for zero-shot 3D semantic segmentation and NuScenes-QA~\cite{qian2024nuscenesqa} for VQA benchmarks. Additionally, DriveVLM~\cite{tian2024drivevlm} employs chain-of-thought reasoning for scene analysis, while DriveDreamer~\cite{wang2025drivedreamer} predicts future states for proactive responses.

\subsubsection{Decision-Making and Control}
\label{sec:control}

FMs improve decision-making and control by translating scene understanding into safe actions. LLMs in Drive as You Speak~\cite{cui2024drive} and LanguageMPC~\cite{sha2023languagempc} process complex data for real-time decisions. Hybrid systems like BEVGPT~\cite{wang2024bevgpt} and Driving with LLM~\cite{chen2024driving} combine reasoning with traditional controls, while SurrealDriver~\cite{jin2024surrealdriver} and Drive Like a Human~\cite{fu2024drive} enhance robustness through safety and memory modules.

\subsubsection{Navigation and Planning}
\label{sec:planning}

FMs integrate natural language into navigation and planning by converting textual instructions into spatial representations. Systems like Talk to the Vehicle~\cite{sriram2019talk} and Ground then Navigate~\cite{jain2023ground} generate waypoints and trajectories from multi-modal inputs. ALT-Pilot~\cite{omama2023altpilot} enhances planning with language-augmented maps using CLIP~\cite{radford2021learning}, while GPT-Driver~\cite{mao2023gptdriver} and DriveVLM~\cite{tian2024drivevlm} support predictive planning and reasoning.

\subsubsection{End-to-End Autonomous Driving}

Recent advancements in FMs have enabled the development of unified models that integrate perception, reasoning, and control into a single differentiable framework. DriveGPT4~\cite{xu2024DriveGPT4} processes sensor inputs and queries for control signals and explanations. ADAPT~\cite{jin2023adapt} maps video to actions and narratives, DriveMLM~\cite{cui2025drivemlm} integrates LLMs into closed-loop systems, and VLP~\cite{pan2024vlp} promotes generalization with context-aware frameworks.

\subsection{Foundation Models for Autonomous Driving System Development}\label{fm4developemnt}
Besides the direct integration into autonomous driving systems, FMs have also been adopted in the development (i.e., testing) of autonomous driving systems. 

\subsubsection{Critical Scenario Generation and Comprehension} Due to its extraordinary capability in understanding the diverse driving environment and generating codes, FMs have been widely used in generating critical scenarios to test the autonomous driving systems~\cite{song2025generative, petrovic2024llm}. For instance, Tang~\textit{et al.}~\cite{tang2024legdend} propose a top-down fashion approach to generate diverse critical scenarios, which utilizes two LLMs to transform functional scenarios to formal scenarios and then searches with the logical scenarios for critical scenarios. Zhang \textit{et al.}~\cite{zhang2024chatscene} propose ChatScene, a LLM-based agent that can generate domain-specific code from text instructions/descriptions, which can then be used to construct the desired scenario in simulators. While significant research effort has been made, existing approaches may still have problems in areas like long-tail scenario generation or multi-modal scenario fusion.

\subsubsection{FM-Based Code Generation} As LLMs have been widely adopted in the daily code writing process, it is inevitable that autonomous driving developers use LLMs to generate code for production use. However, as pointed out by \cite{jimenez2024swebench, liu2023is}, LLMs can generate erroneous code, or even worse, code that can pass the unit test but contain potential vulnerabilities. Nouri~\textit{et al.}~\cite{nouri2025simulation} propose a prototype for automatic code generation and assessment using a designed pipeline of an LLM-based agent, simulation model, and a rule-based feedback generator. This pipeline can automatically evaluate the LLM-generated code and generate an assessment report as feedback to the LLM for modification or bug-fixing, which is a valuable attempt. But considering the ADS is a safety-critical software system, significant research effort still needs to be made.

\subsection{Review Methodology}
We conducted a structured literature review to identify representative studies on adopting FM in ADSs and to support the synthesis and roadmap proposed in this paper. Our objective is to provide a transparent and reproducible account of how the literature was collected and organized, rather than to claim exhaustive coverage of all publications in this rapidly evolving area.

\subsubsection{Scope and time span} We focus on the period from January 2020 to January 2026, which covers the emergence and rapid adoption of large-scale pre-trained (LLMs, VLMs, and world models) and their application to AD-related tasks. We considered works that address autonomous driving tasks, FM infrastructure, in-vehicle integration, real-world deployment, and engineering practices.

\subsubsection{Database Search} We identified candidate papers through scholarly database search. In particular, we select \emph{DBLP}\footnote{\url{https://dblp.org/}} as our database, which is a popular bibliography database containing a comprehensive list of research venues in computer science. Furthermore, this initial search targets the titles of the papers, as the title often conveys the theme of a paper~\cite{tang2023survey}. The search string is optimized in an iterative manner to cover as many related papers as possible. The final search string is shown as follows:

\smallskip
\setlength{\fboxsep}{8pt}
\begin{center}
\Ovalbox{
\begin{minipage}{0.9\textwidth}
\small ((``foundation model'' OR ``large language model'' OR ``vision language model'' OR ``large vision language model'' OR ``world model'' OR ``multimodal model'' OR ``FM'' OR ``LLM'' OR ``VLM'' OR ``MLLM'') \\
AND \\
(``automated vehicle'' OR ``automated driving'' OR ``autonomous car'' OR ``autonomous vehicle'' OR ``autonomous driving'' OR ``self-driving'' OR ``driver assistance system'' OR ``intelligent vehicle'' OR ``intelligent agent''))
\end{minipage}
}
\end{center}
\smallskip

The first group of terms (above ``AND'') represents the identified synonyms of foundation models, containing the terms such as ``large language model'' and ``foundation model''; The second group of terms (below ``AND'') contains the synonyms of autonomous driving, containing the terms such as ``autonomous driving'' and ``self-driving''. The terms in each group are connected with~\emph{OR} operator, while the two groups are connected using~\emph{AND} operators, which ensures that the results should contain the characteristics of both groups. Additionally, we observed that certain topics (e.g., hallucination, foundation model alignment) have a limited number of publications related to autonomous driving. For instance, the search string (``autonomous driving'' AND ``hallucination'') only returns two results. Thus, for specific topics related to FMs (e.g., ``hallucination''), we substitute the second group in the search string with the corresponding \emph{``<topic>''} to gain a more comprehensive understanding.

\subsubsection{Abstract Analysis}To determine whether a paper is relevant to the scope of this paper, we manually screened the title and abstract of each candidate paper according to the inclusion (\textbf{IC\#}) and exclusion (\textbf{EC\#}) criteria below. When the relevance was unclear from the abstract, we further inspected the full text.

\noindent Inclusion Criteria:
\begin{itemize}
    \item \textbf{IC1:} propose or evaluate FMs for autonomous driving tasks;
    \item \textbf{IC2:} introduce datasets, benchmarks, toolchains, or evaluation protocols that are used to study FM-based ADS capabilities;
    \item \textbf{IC3:} discuss deployment- and system-oriented issues for FM-based ADSs, such as latency, safety assurance.
    \item \textbf{IC4:} papers published between January 2020 and January 2026.
\end{itemize}

\noindent Exclusion Criteria:

\begin{itemize}
    \item \textbf{EC1:} preprint papers or non-peer-reviewed papers (except for technical reports from leading companies or dataset/benchmarks);
    \item \textbf{EC2:} papers that are not related to ADS (e.g., general robotics, unmanned aerial vehicle);
    \item \textbf{EC3:} papers provide insufficient technical details or early results.
\end{itemize}

Specifically, we retain a small number of non-peer-reviewed arXiv papers when they serve as primary technical reports from leading industrial labs or companies and document influential FM systems, datasets, benchmarks, or deployment practices that are not yet available in peer-reviewed venues; we do not exclude survey or roadmap papers, which we instead record and compare in~\Cref{sec:related}; for \textbf{EC2}, papers related to other intelligent systems, such as unmanned aerial vehicles, are excluded, since we focus on autonomous driving systems.

\subsubsection{Limitations} Given the fast pace of FM research, the literature is continuously expanding. While we aimed for broad coverage, we do not claim exhaustiveness. Our focus is on representative and influential works that enable a structured synthesis and support actionable research directions for the community.

\subsection{Comparison with Related Surveys and Roadmaps}\label{sec:related}
We would like to clarify the difference between our roadmap and previous related surveys (or roadmaps)~\cite{song2025generative},~\cite{cui2024survey},~\cite{yan2024forging},\\~\cite{gao2025survey},~\cite{zhou2024vision},~\cite{xie2025are},~\cite{wang2025generative},~\cite{li2024large},~\cite{wu2025multi},~\cite{zhu2025survey},~\cite{yang2024llmdrive},~\cite{guan2024world},~\cite{tu2025role}~\cite{wu2024prospective}~\cite{gao2025foundation},~\cite{dai2025large},~\cite{sathyam2025foundation},~\cite{jiang2025survey},~\cite{feng2025survey},~\cite{tian2026large},~\cite{dong2025end}. Many prior works primarily focus on discussing how ADS can be empowered by FMs, and how FMs may reshape the autonomous driving. Among these works, Zhou~\emph{et al.}~\cite{zhou2024vision} provide a comprehensive review of advances in VLMs for empowering autonomous driving tasks and summarize several challenges, and Cui~\emph{et al.}~\cite{cui2024survey} presented a survey of the application of MLLMs in autonomous driving tasks. Li~\emph{et al.}~\cite{li2024large} and Zhu~\emph{et al.}~\cite{zhu2025survey} both survey LLM-enabled autonomous driving, organizing applications across the autonomous driving tasks and highlighting key challenges for safe deployment (e.g., real-time constraints, hallucination), and Yang~\emph{et al.}~\cite{yang2024llmdrive} summarize the research landscape regarding the application of LLMs and VLMs in autonomous driving. Xie~\emph{et al.}~\cite{xie2025are} conduct an empirical study evaluating the readiness of VLMs for autonomous driving, highlighting their strengths in semantic scenario understanding while identifying critical weaknesses in spatial reasoning and long-horizon planning, while Dong~\emph{et al.}~\cite{dong2025end} survey end-to-end autonomous driving, tracing its evolution from classic imitation learning (IL)/reinforcement learning (RL) to FM-empowered approaches and summarizing key challenges and future directions. Gao~\emph{et al.}~\cite{gao2025survey} survey foundation models for autonomous driving, proposing a modality-and-function taxonomy (LLMs, vision FMs, and multimodal FMs) and summarizing their roles in planning, perception, prediction, and simulation alongside key deployment limitations (e.g., hallucination and efficiency), Jiang~\emph{et al.}~\cite{jiang2025survey} survey vision-language-action (VLA) models for autonomous driving, unifying architectures, representative systems, datasets/benchmarks, and evaluation protocols, and outlining open challenges such as robustness, real-time efficiency, and formal verification, and Tian~\emph{et al.}~\cite{tian2026large} survey LLM/VLM integration for autonomous vehicles across modular and end-to-end pipelines, data generation, and evaluation resources, highlighting practical deployment issues such as real-time efficiency and ethical/regulatory concerns. Wu~\emph{et al.}~\cite{wu2024prospective} provide a comprehensive synthesis of the roles various FMs play in advancing autonomous driving safety by enhancing various autonomous driving tasks, and facilitating data augmentation to address long-tail distribution challenges, and Yan~\emph{et al.}~\cite{yan2024forging} provide a systematic survey of over 250 papers to categorize the evolution of vision foundation models, detailing advancements in data curation, pre-training strategies, and downstream adaptation for autonomous vehicles. Wu~\emph{et al.}~\cite{wu2025multi} investigate the integration of LLMs in multi-agent ADS, and Dai~\emph{et al.}~\cite{dai2025large} study the use of FMs for trajectory prediction in autonomous driving, while Sathyam~\emph{et al.}~\cite{sathyam2025foundation} examine their use for perception. Furthermore, Guan~\emph{et al.}~\cite{guan2024world}, Tu~\emph{et al.}~\cite{tu2025role}, and Feng~\emph{et al.}~\cite{feng2025survey} conduct comprehensive reviews of the role of world models in autonomous driving tasks. Wang~\emph{et al.}~\cite{wang2025generative} delivered a comprehensive synthesis of generative AI's role in autonomous driving, mapping foundational architectures like LLMs to frontier applications in multimodal data generation, simulation, and reasoning, while evaluating the technical and ethical challenges, Song~\emph{et al.}~\cite{song2025generative} surveys the application of generative AI in testing autonomous driving systems, and Gao~\emph{et al.}~\cite{gao2025foundation} survey how FMs enable driving scenario generation and scenario analysis for scenario-based testing, providing a unified taxonomy of methods, datasets/simulators, metrics, and open challenges.

While some challenges discussed in this paper have already been proposed in the surveys listed above, for instance, hallucination by~\cite{li2024large, zhu2025survey} and real-time constraints by~\cite{tian2024drivevlm, li2024large}, they generally only point out these challenges without in-depth investigation. Although Wu~\emph{et al.}~\cite{wu2024prospective} have conducted exploratory experiments using quantization techniques to showcase the possibility of deploying LLM-based ADS in practice, their experiments remain illustrative and do not consider the potential impact of quantizing LLMs on ADS tasks. In contrast, our work covers broader topics across the three dimensions of integrating FMs into autonomous driving systems and provides in-depth discussions. In particular, we approach these challenges through the lens of software engineering. For each challenge, we first discuss its current status and potential impact, then we summarize existing SE approaches to mitigate it and their limitations, and finally point out the opportunities.

\subsection{Overview of the Research Landscape}
To provide a structured overview of the research landscape, this section summarizes representative works identified through our systematic review, including their tasks, evaluation metrics, and key limitations. The tables also include representative background artifacts (e.g., widely used frameworks, industrial products) that are outside the primary DBLP review corpus but are included to situate the infrastructure and deployment landscape. We categorize the landscape into two primary dimensions: the underlying \textbf{Infrastructure} (Table~\ref{tab:infrastructure_landscape}), which covers the data, deployment toolchains, and hardware roots of trust required to support large-scale models, and the \textbf{In-Vehicle Application} (Table~\ref{tab:fm_in_vehicle_landscape}), which details how FMs are integrated into core ADS modules such as perception, planning, and control.

Table~\ref{tab:infrastructure_landscape} highlights a critical shift in the field; while classical ADS research focused on modular performance, the integration of FMs introduces complex software engineering challenges such as GPU memory saturation during training and the need for TEE-based isolation for high-bandwidth sensor data. Simultaneously, Table~\ref{tab:fm_in_vehicle_landscape} illustrates the transition toward end-to-end multi-modal models (e.g., DriveVLM~\cite{tian2024drivevlm} and DriveGPT4~\cite{xu2024DriveGPT4}), which aim to bypass brittle perception-to-planning cascades but face new hurdles in real-time latency and "hallucination" precision.

While the works presented in these tables are selected for their representative value in illustrating these SE challenges, they are by no means exhaustive. For readers seeking a more exhaustive list of foundation models and their specific architectural configurations in the broader AI domain, we recommend referring to recent comprehensive surveys such as those by Tian~\textit{et al.}~\cite{tian2026large} and Zhu~\textit{et al.}~\cite{zhu2025survey}, which provide extensive taxonomies of individual model variants across the AD ecosystem.

\begin{landscape}
\begin{table}[p] 
    \small
    \renewcommand\arraystretch{1.5} 
    \setlength\tabcolsep{6pt}
    \caption{Infrastructure Landscape for Integrating Foundation Models into Autonomous Driving Systems}
    \label{tab:infrastructure_landscape}
    
    \begin{tabularx}{\linewidth}{
        >{\hsize=0.6\hsize}X 
        >{\hsize=0.7\hsize}X 
        >{\hsize=1.5\hsize}X 
        >{\hsize=1.1\hsize}X 
        >{\hsize=1.1\hsize}X 
    }
    \toprule
    \textbf{Category} & \textbf{Task} & \textbf{Representative Work (Product)} & \textbf{Key Metrics} & \textbf{Limitations} \\ \midrule

    \multirow{5.5}{=}{Dataset} 
        & Augmentation & Automated annotation approaches~\cite{arai2025covla, chen2024panda}, critical scenario generation~\cite{zhang2024chatscene, ding2020learning, wang2021advsim}, scenario reconstruction~\cite{tang2024legdend, wei2024editable}, scenario transformation~\cite{baresi2025dillema}  & Caption accuracy, diversity, reconstruction consistency, multi-modal alignment & Sim2Real gap; hallucination in VLM annotated datasets; multi-modality alignment; diversity;  \\ \cmidrule{2-5}
        & Dataset Debiasing & Fairness testing~\cite{li2024bias} and evaluation~\cite{ryu2024designing, khoshkdahan2025fair, khoshkdahan2025overall}, mitigation~\cite{llorca2024attribute, amini2018variational, katare2025analyzing, aylapuram2025bias} & Statistical Parity Difference (SPD), Equal Opportunity Difference (EOD), Average Odds Difference (AOD), Intersection over Union (IoU) Disparity, Detection Miss Rate (MR) (for pedestrian detection)~\cite{chen2023comprehensive, li2024bias} & Primarily focus on perception module (e.g., pedestrian detection), limited investigation about its impact to the decision; integrating FMs may introduce inherited bias which needs further inspection.\\ \cmidrule{2-5}
        & Privacy Preserving & Privacy-preserving machine learning~\cite{zhou2025privacy,zeng2025epauto}, federated learning~\cite{xiang2025flad, yang2025fedgpl}  & accuracy, communication overhead, client selection rate, data heterogeneity, differential privacy budgets & High communication overhead for FM models; FMs tend to memorize training data and can even infer personal attributes from anonymized data~\cite{staab2024beyond}.  \\ \midrule

    \multirow{6}{=}{Development \& Deployment} 
        & Development Framework & PyTorch~\cite{paszke2019pytorch}, TensorFlow~\cite{tensorflow2015whitepaper}, JAX~\cite{jax2018github}, PaddlePaddle~\cite{bi2022paddle} & Training time, accuracy, inference time, CPU/GPU utilization, memory usage, community support & Only offer basic distributed training functions, and exhibit critical limitations when scaling to FM-level workloads (e.g., GPU memory saturation, communication bottlenecks)~\cite{liu2023rise, yu2025understanding}. \\ \cmidrule{2-5}
        & Distributed Training & Ray~\cite{moritz2018ray}, DeepSpeed~\cite{jeff2020deepspeed}, Megatron-LM~\cite{megatron-lm}, Colossal-AI~\cite{li2023colossal} & Scaling efficiency, training throughput, memory efficiency, stability & High onboarding efforts for developers; limited research on bug comprehension~\cite{yu2025understanding, ma2025comprehensive} and testing techniques~\cite{wang2025d3}. \\ \cmidrule{2-5}
        & Efficient Deployment & Model pruning~\cite{ilhan2024resource,li2025owled}, model distillation~\cite{shaw2025cleanmap,zhong2025clip4retrofit, hedge2025distilling}, and quantization~\cite{uprety2025edge, lin2024awq} & Inference latency, memory footprint, accuracy retention & Performance drop; limited investigation on real-world testing and its impact on robustness~\cite{sudharshana2025comprehensive}. \\ \midrule

    \multirow{4}{=}{Computational Resources} 
        & Hardware & AVA-3510 (based on NVIDIA Quadro RTX 5000)~\cite{adlink_ava_3510}, ADM-AL30 (based on NVIDIA RTX 4000 SFF Ada)~\cite{adlink_adm_al30}, NVIDIA DRIVE AGX Thor~\cite{nvidia_thor}, Tesla FSD Hardware (custom-designed SoC)~\cite{talpes2020compute} & TFLOPS, Memory Bandwidth, GPU Memory Size & Black-box proprietary designs are hard to verify; only a few products (e.g., AGX Thor) conform to safety standards like ISO 26262 ASIL-D~\cite{iso26262_part9}. \\ \cmidrule{2-5}
        & TEE for AD & CVShield~\cite{hu2020cvshield}, Mimer Trust~\cite{cai2022mimer} & Attestation speed, Enclave size, overhead & Current approaches are primarily limited to low-bandwidth sensor channels like GPS; limited evaluation for high-bandwidth sensors (LiDAR/Camera); Limited enclave size for FM-level model; \\ \bottomrule
    \end{tabularx}
\end{table}
\end{landscape}

\begin{table}[htbp]
\footnotesize
\renewcommand\arraystretch{1.25}
\caption{FM in Vehicle Landscape}
\label{tab:fm_in_vehicle_landscape}

\begin{tabularx}{\textwidth}{
@{}
>{\raggedright\arraybackslash}p{2.35cm}   
>{\hsize=0.95\hsize\raggedright\arraybackslash}X 
>{\hsize=0.70\hsize\raggedright\arraybackslash}X 
>{\hsize=0.75\hsize\raggedright\arraybackslash}X 
>{\hsize=1.05\hsize\raggedright\arraybackslash}X 
>{\hsize=1.55\hsize\raggedright\arraybackslash}X 
@{}
}
\toprule
\textbf{Task} & \textbf{Paper} & \textbf{Model} & \textbf{Dataset/Sim} & \textbf{Key Metrics} & \textbf{Limitations} \\
\midrule

\multirow[t]{3}{*}{\makecell[l]{\textbf{Perception}\\\textbf{and Understanding}}}
& Semantic Anomaly Detection with LLMs~\cite{elhafsi2023semantic}
& GPT-3.5~\cite{openai2023chatgpt35} & CARLA~\cite{carla}
& TPR, FPR
& \multirow[t]{3}{=}{Strong coupling with upstream perception; hallucination; poor spatial precision; temporal inconsistency;} \\
\cmidrule(lr){2-5}
& Zelda~\cite{romero2023zeldavideoanalyticsusing}
& VIVA~\cite{romero2022optimizing}
& BDD-X~\cite{kim2018textual}
& MAP, APS
& \\
\cmidrule(lr){2-5}
& Talk2BEV~\cite{choudhary2024Talk2BEV}
& BLIP-2~\cite{li2023blip2}, MiniGPT-4~\cite{zhu2024minigpt}, InstructBLIP~\cite{dai2023instructblip}
& Talk2BEV-Bench~\cite{choudhary2024Talk2BEV}
& MCQ Acc., IoU, Distance error
& \\
\midrule

\multirow[t]{3}{*}{\makecell[l]{\textbf{Decision Making}\\\textbf{and Control}}}
& LanguageMPC~\cite{sha2023languagempc}
& GPT-3.5 & IdSim~\cite{jiang2024reinforcement}
& Failure and collision cases, inefficiency, time efficiency, safety penalty
& \multirow[t]{3}{=}{High on-device latency; struggles with long-horizon goals; high sensitivity to minor prompt disruptions; potential inheritance of bias from VLM/LLMs~\cite{zhu2025survey};} \\
\cmidrule(lr){2-5}
& Driving with LLMs~\cite{chen2024driving}
& LLaMA-7b~\cite{touvron2023llama} & Custom 2D Simulator
& Mean Absolute Error (MAE) for perception and action; traffic light detection accuracy; GPT/Human Grading for QA 
& \\
\cmidrule(lr){2-5}
& Drive Like a Human~\cite{fu2024drive}
& GPT-3.5 (with LLaMA-Adapter) & HighwayEnv~\cite{leurent2018highwayenv}
& Zero-shot pass rate, reasoning accuracy, and decision consistency
& \\
\midrule

\multirow[t]{3}{*}{\makecell[l]{\textbf{Navigation}\\\textbf{and Planning}}}
& ALT-Pilot~\cite{omama2023altpilot}
& GPT-4, CLIP
& CARLA + field test
& Absolute Position Error (APE), Recall@K, Distance to Closest Landmark Region (DCLR), Distance to Converge, and Goal Reachability
& \multirow[t]{3}{=}{Risk of generating unaligned/dangerous planning/routines; poor multi-turn interaction and instruction following;} \\
\cmidrule(lr){2-5}
& GPT-Driver~\cite{mao2023gptdriver}
& GPT-3.5 & nuScenes~\cite{caesar2020nuscenes}
& Avg. L2 Error (m) and Avg. Collision Rate
& \\
\cmidrule(lr){2-5}
& DriveVLM~\cite{tian2024drivevlm}
& Qwen-VL~\cite{qwen-vl}
& SUP-AD~\cite{tian2024drivevlm}, nuScenes
& Displacement Error (DE), Collision Rate (CR), Scene Description Score, and Meta-action Score
& \\
\midrule

\multirow[t]{3}{*}{\makecell[l]{\textbf{End-to-End}\\\textbf{Autonomous Driving}}}
& DriveGPT4~\cite{xu2024DriveGPT4}
& GPT-4 & BDD-X
& CIDEr, BLEU4, ROUGE-L, ChatGPT Score, RMSE, and Threshold Accuracies ($A_\tau$)
& \multirow[t]{3}{=}{High \textbf{on-device latency}; risk of executing unaligned/dangerous user commands; transparency and interpretability;} \\
\cmidrule(lr){2-5}
& DriveMLM~\cite{cui2025drivemlm}
& LLaMA2-7B, EVA-CLIP
& CARLA
& Accuracy and F1-measure for decision prediction; BLEU-4, CIDEr, and METEOR for decision explanation; L2 distance, collision rate, and intersection violation rate for trajectory prediction
& \\
\cmidrule(lr){2-5}
& VLP~\cite{pan2024vlp}
& CLIP~\cite{radford2021learning}
& nuScenes
& Avg. L2 error, and Avg. collision rate
& \\
\bottomrule
\end{tabularx}
\end{table}

\section{FM4AD Infrastructure}
\label{sec:infrastructure}

The infrastructure is the cornerstone for integrating FMs into autonomous driving, encompassing the datasets, computational resources, and toolchains necessary for training, testing, and deployment of FMs.

\subsection{High-quality Dataset for Autonomous Driving}\label{sec:dataset}
High-quality datasets have played a critical role in advancing autonomous driving technology. Traditional datasets primarily focused on 2D annotations like bounding boxes and masks for RGB camera images~\cite{pan2018spatial}. With the emergence of FMs, datasets are evolving towards multi-modal integration, particularly incorporating language descriptions~\cite{zhou2024vision}. While this multi-modal approach promises to accelerate autonomous driving development, it introduces new challenges. Moreover, the massive data requirements for training FMs raise significant concerns about privacy protection and ethical/legal compliance~\cite{longpre2024large}. Thus, we identify the following challenges:

\noindent \textbf{Challenge I: Dataset Cleaning and Curation.} Dataset cleaning and curation serve as a critical foundation for developing FMs, ensuring data integrity, privacy protection, and efficient training. Key challenges include protecting privacy~\cite{zhu2024privauditor, zhang2024towards, carlini2021extracting}, and mitigating bias in training datasets~\cite{llorca2024attribute, li2024bias, hort2024bias}. The privacy challenge involves both preventing personal data from appearing in training datasets and ensuring it cannot be inferred from model outputs~\cite{kim2023propile, zhu2024privauditor}. Bias in training data can lead models to perpetuate systemic inequities and may introduce safety risks for unrepresented groups when deployed in real-world scenarios~\cite{llorca2024attribute}. These challenges present several key opportunities for further research.

\begin{itemize}[leftmargin=15pt]
    \item \textbf{Opportunity: Bias Mitigation.} Dataset bias poses a critical challenge for autonomous driving. Recent studies~\cite{llorca2024attribute} have revealed limited diversity in key demographic attributes (i.e., age, sex, and skin tone) within existing AD datasets. This lack of representation could lead to safety risks when deploying FMs trained on such datasets, particularly for underrepresented groups. While recent research has made progress in addressing these concerns~\cite{li2024bias, hort2024bias, hort2024search, chen2025diversity}, these previous works mainly focus on the perception module (i.e., pedestrian detection), using fairness metrics such as Statistical Parity Differences (SPD), Equal Opportunity Difference (EOD), Average Odds Difference (AOD), and Miss Rate (MR). One open software engineering challenge lies in verifying how such biases propagate through the system pipeline. Future research could develop stratified, scenario-based testing suites where only demographic attributes vary, and evaluate whether downstream behaviors (e.g., safety margin, rule violations, intervention rate) exhibit systematic disparities, enabling fairness regression testing across releases. In addition, LLMs are known to exhibit bias and fairness issues~\cite{gallegos2024bias}; integrating FMs into ADS reasoning/planning may introduce new bias pathways beyond perception, which calls for dedicated test suites and systematic investigation.
    \item \textbf{Opportunity: Privacy Preservation.} Another critical research opportunity lies in developing effective privacy-preserving algorithms for FMs in autonomous driving. According to Staab \emph{et al.}~\cite{staab2024beyond}, LLMs can even infer personal attributes from real-world data, even when the text is anonymized using commercial tools. Wang~\emph{et al.}~\cite{wang2026secure} highlight that malicious attackers can infer and trace vehicle trajectories in accident warning systems, raising serious privacy concerns and introducing potential security risks. Established techniques such as differential privacy~\cite{duan2023flocks,chua2024mind,tang2024privacypreserving,tramer2024position}, data cleaning~\cite{kandpal2022deduplicating,carlini2021extracting}, and federated learning~\cite{xu2024fwdllm, zhang2024towards, sun2024fedbpt},  have advanced the field, yet they consistently face challenges in balancing privacy preservation with data utility. The challenge is particularly acute with FMs, which tend to memorize training data extensively, potentially leading to privacy leakage even with the data used in fine-tuning processes~\cite{zhu2024privauditor}. There is an urgent need for novel techniques that can ensure robust privacy guarantees while maintaining the comprehensive nature of training datasets required for large FMs.
    \item \textbf{Opportunity: Machine Unlearning for FMs.} While curating training data can mitigate issues upfront, FMs already deployed may contain private information or learned biases. For these models, machine unlearning emerges as a critical capability~\cite{xu2023machine}. The goal is to efficiently remove the influence of specific data points or concepts from a trained model without the need for a complete, costly retraining from scratch. This is essential for complying with data privacy regulations like the "right to be forgotten" and for rectifying harmful biases discovered post-deployment~\cite{xu2023machine}. For coding tasks, machine unlearning has proven its effectiveness~\cite{chu2025scrub}. The challenge of this opportunity lies in developing unlearning techniques that are not only effective at removing information but are also computationally efficient and maintain the model's overall performance on other tasks, which is particularly difficult for FMs in autonomous driving, since these models inherently need to deal with multiple tasks.
\end{itemize}

\noindent \textbf{Challenge II: Augmenting Autonomous Driving Datasets.} Despite significant investments in the development of autonomous driving datasets, current limitations in data quality and scale hinder their ability to comprehensively address the field’s challenges~\cite{cui2024survey}. Moreover, certain critical scenarios remain difficult or nearly impossible to capture in real-world data collection~\cite{wei2024editable, song2024synthetic}. These include high-risk situations such as accident aftermath and pedestrian-involved incidents. However, comprehensive testing of autonomous vehicles against these scenarios is crucial for safety validation. To overcome these challenges, researchers should increasingly explore methods for generating customized driving scenarios or automated data collection, enabling the effective simulation of these critical cases to augment existing datasets and enhance their utility.

\begin{itemize}[leftmargin=15pt]
    \item \textbf{Opportunity: Customizable Driving Scenario Generation.} Current autonomous driving systems are primarily trained and evaluated on datasets collected from daily driving scenarios or synthetic data~\cite{ding2023survey, wei2024editable}. However, these datasets generally lack safety-critical scenarios that are crucial for robust system evaluation. Research in driving scenario generation has progressed along multiple directions, including data-driven approaches~\cite{kruber2019unsupervised,tan2021scenegen,pronovost2023scenario}, adversarial generation methods~\cite{ding2020learning,wang2021advsim, Abeysirigoonawardena2019generating, mei2025llmattacker}, and knowledge-based techniques~\cite{zhang2024chatscene, tang2024legdend, wei2024editable}. Looking ahead, scenario generation algorithms need to address key challenges, such as maintaining consistency across multiple sensor modalities (e.g., LiDAR, camera images) and enhancing scenario complexity through interaction and collaboration between agents.
\end{itemize}

\noindent \textbf{Challenge III: Dataset Licensing and Management.} Dataset licensing and management pose a variety of challenges vital to ensuring the legal and ethical use of autonomous driving datasets. The massive amount of data required for training FMs heightens the risks of copyright breaches, licensing violations, and subsequent legal liabilities. Additionally, the terms of use for datasets released by leading autonomous driving companies vary widely, further complicating this task. The multimodal nature and diverse sources of autonomous driving datasets intensify these difficulties. Moreover, selecting/sampling the right training data is essential for producing capable FMs~\cite{wettig2024qurating, brown2020lm, chowdhery2023palm}. Recent studies~\cite{wolter2023open, kim2024donottrust} have revealed the complex landscape of modern large dataset copyright and licensing, emphasizing the need for deeper exploration and development of innovative techniques. These challenges also open up opportunities for further research.

\begin{itemize}[leftmargin=15pt]
    \item \textbf{Opportunity: Dataset License Compliance.} The primary challenge of license management lies in the complexity and variety of licenses governing autonomous driving datasets~\cite{jahanshahi2025cracks, xu2025evaluating}. Unlike traditional datasets for LLMs, which primarily consist of publicly available data (crawled from the Internet) supplemented with proprietary datasets having usage restrictions, most autonomous driving datasets are released by leading autonomous driving companies with their own specific terms of use, necessitating careful review and understanding to ensure compliance~\cite{wolter2023open, lucchi2024chatgpt,vendome2017machine}. As pointed out by Kim \emph{et al.}~\cite{kim2024donottrust}, the scale of modern datasets renders manual compliance verification impractical, thereby requiring automated detection techniques. Promising research directions include the development of automated detection and audit systems for legal terms of use, providing developers with clear insights into the permissions and restrictions associated with each dataset.

    \item \textbf{Opportunity: Data Management Framework for FMs.} As FMs demonstrate performance improvements through data scaling and the significance of data becomes evident~\cite{chowdhery2023palm}, effective data management becomes increasingly critical. While various tools and methods have been proposed to explore how to properly manage the training data, encompassing data deduplication~\cite{lee2022deduplicating}, training data selection~\cite{rae2022scaling, wang2024greats, li2024quantity}, sampling high-quality data~\cite{peng2025dataman,wettig2024qurating}, and dataset license compliance~\cite{kim2024donottrust, jahanshahi2025cracks}, there is still a lack of a unified framework and criteria. Although there have been some initial attempts in this area~\cite{wang2024data,ostendorff2024llmdatasets, peng2025dataman}, systematic approaches to data management for FMs remain in their early stages. Given the massive scale and diverse sources of data required for training FMs, developing a comprehensive data management framework has become an urgent priority.
\end{itemize}

\subsection{Computational Resources}\label{sec:comutational_resource}
\label{sec:computational}
Due to the computational-intensive nature of FMs, computational resources, including graphics processing units (GPUs), tensor processing units (TPUs), and other specialized AI accelerators, form the very foundation of the FM infrastructure. Building upon this hardware layer, distributed training frameworks and cloud computing enable efficient resource utilization and management. However, the complexity of distributed, computation-intensive training, and reliable efficient deployment introduces unique challenges and opportunities in adopting FMs in autonomous driving. In this section, we mainly discuss the challenges and opportunities related to hardware, issues with the software layer (e.g., distributed training framework) are discussed in Section~\ref{sec:dev_deploy}.

\noindent \textbf{Challenge I: Trustworthy Hardware Design.} The computationally intensive nature of FMs necessitates a reliance on proprietary and specialized hardware, such as GPUs and other accelerators. This dependency creates significant challenges in ensuring security across the hardware stack, as the proprietary design of chips and firmware often results in a "black-box" environment where vulnerabilities can remain undetected~\cite{wang2024large, yang2016a2}. Furthermore, the parallel processing architectures and shared resources inherent to modern accelerators make them particularly susceptible to hardware-level attacks~\cite{jiang2017novel}. These attacks can lead to severe consequences, including the leakage of sensitive information like model parameters via side-channels~\cite{Naghibijouybari2021side} or even enabling arbitrary code execution~\cite{lee2022securing}. Given the paramount importance of security in autonomous driving, these hardware vulnerabilities present a fundamental risk that necessitates urgent research and development.
Beyond the general exposure of proprietary accelerators, FM-centric ADS deployments introduce additional constraints that exacerbate the hardware trust problem. First, FM inference (and especially VLM/world-model pipelines) typically requires large model footprints (weights, intermediate activations, and KV caches), which can exceed the secure memory/enclave capacity of many practical TEE designs and can trigger expensive enclave paging and frequent world switches. Second, ADS workloads ingest multi-channel, high-bandwidth sensor streams (e.g., camera and LiDAR), so naïvely confining the full perception--planning pipeline to a secure enclave can incur substantial I/O marshalling overhead and may violate real-time latency budgets. Existing TEE-for-AD prototypes are often evaluated on low-bandwidth channels (e.g., GPS) rather than on realistic high-throughput perception inputs~\cite{hu2020cvshield}, leaving the end-to-end feasibility for FM-scale workloads under-explored. Third, in today’s deployment reality, widely used commodity accelerators may not expose confidential-computing capabilities suitable for isolating FM execution, while platforms that do provide stronger TEE-style protections can be substantially more costly and harder to provision at scale; this gap motivates practical designs that minimize the trusted computing base (TCB) and carefully trade off security guarantees against latency and integration overhead.

\begin{itemize}[leftmargin=15pt]
     \item \textbf{Opportunity: Security-by-Design Hardware Architectures.} A primary research opportunity lies in architecting hardware with intrinsic security guarantees, with a key approach being the development of Trusted Execution Environments (TEEs) for AI accelerators~\cite{yao2024survey, chen2024llm}. A TEE~\cite{chen2020training} would leverage hardware to create an isolated enclave, protecting the confidentiality and integrity of an FM's parameters and its execution, even from a compromised high-level operating system.
    From a deployment perspective, widely used commodity accelerator stacks can provide limited support for enclave-style protection. Prior work reports that commonly used commercial GPUs in current AV/AD solutions/platforms (e.g., RTX A5000 in AVA-3510) do not provide TEE functionality, whereas confidential-computing-enabled accelerators (e.g., Hopper-class GPUs) can be substantially more costly~\cite{li2025teeslice}. This deployment gap motivates hybrid designs that minimize the trusted computing base (TCB) and confine only security-critical components to the trusted domain.
    Concretely, TEEs for FM-powered ADSs must satisfy architectural requirements such as (i) \emph{low overhead and real-time predictability} (bounded enclave transitions and secure I/O costs), (ii) \emph{scalability} to FM-scale working sets (or secure partitioning/streaming of model components), (iii) \emph{accelerator-aware trust boundaries} that account for GPU/NPU driver stacks and DMA paths that often remain outside the enclave, and (iv) \emph{multi-sensor support} for securely ingesting and attesting to high-throughput perception data.
    Significant research is therefore needed to design novel, low-overhead TEEs that can scale to serve FMs, establishing a verifiable hardware root of trust for critical AI computations in autonomous driving.
    To make the comparison space explicit, as shown in~\Cref{tab:tee_categories}, TEEs can be classified along two orthogonal axes---\emph{software- vs.\ hardware-based} and \emph{privileged vs.\ non-privileged} deployments---each implying different isolation granularity, attestation support, and performance costs~\cite{munoz2023survey}. A systematic evaluation of these design points for FM workloads (secure memory limits, sensor I/O throughput, and accelerator integration) remains an open problem and is essential for identifying feasible, automotive-grade TEE configurations.

    \item \textbf{Opportunity: Open and Verifiable Hardware Stacks.} To address the "black-box" nature of proprietary hardware, there is an opportunity to develop and promote open-source, verifiable hardware designs or standards for FM acceleration. An open and transparent hardware stack would allow for community-driven security audits, reducing the risk of hidden backdoors or design flaws~\cite{wang2024large}. 
    In addition to openness, an important opportunity is to standardize \emph{measurement and attestation interfaces} (e.g., for firmware provenance, model binaries, and critical runtime configurations) so that safety cases can incorporate verifiable evidence about the hardware and software supply chain. Such interfaces would enable reproducible security evaluation across vendors and help bridge the gap between research prototypes and deployable, certifiable FM acceleration stacks.
\end{itemize}

\begin{table*}[t]
    \centering
    \footnotesize
    \renewcommand{\arraystretch}{1.25}
    \setlength{\tabcolsep}{6pt}
    \caption{TEE implementation categories, characteristics, and representative examples (adapted from Muñoz et al.~\cite{munoz2023survey}).}
    \label{tab:tee_categories}
    \begin{tabularx}{\textwidth}{@{}l >{\raggedright\arraybackslash}X >{\raggedright\arraybackslash}X @{}}
        \toprule
        \textbf{TEE Category} &
        \textbf{Description and Characteristics} &
        \textbf{Examples} \\
        \midrule

        \textbf{1) Hardware-based \& Privileged} &
        \textbf{Description:} Rely on dedicated hardware isolation (e.g., ARM TrustZone) and run with high system privileges. \newline
        \textbf{Characteristics:} (i) Access to broad system resources; (ii) typically uses a secure monitor; (iii) two domains: Secure World (SW) and Normal World (NW). &
        \textbf{Commercial:} Qualcomm QSEE; Trustonic t-base; Samsung TZ-RKP; Google Trusty. \newline
        \textbf{Open/Academic:} Linaro OPTEE; Microsoft TLR; SafeG; Kinibi\_M. \\
        \midrule

        \textbf{2) Hardware-based \& Non-privileged} &
        \textbf{Description:} Utilize hardware support (e.g., Intel SGX or AMD SEV) without granting the TEE instance total control over the system. \newline
        \textbf{Characteristics:} (i) Supports multiple deployments (new instances can be added without extending the Trusted Computing Base); (ii) often referred to as \emph{enclaves}. &
        \textbf{Commercial:} Intel SGX; AMD SEV; TrustICE. \newline
        \textbf{Open/Academic:} Sanctum; SecureBlue; Haven; SCONE; Graphene-SGX. \\
        \midrule

        \textbf{3) Software-based \& Privileged} &
        \textbf{Description:} Software-enforced TEEs (without dedicated secure hardware elements) that run with high privileges to enforce isolation. \newline
        \textbf{Characteristics:} (i) Often relies on hypervisor or kernel-level modifications; (ii) isolation is logical rather than physical. &
        \textbf{Open/Academic:} Nested Kernel; OpenTEE; MicroTEE; SoftTEE; TrustShadow; SKEE. \\
        \midrule

        \textbf{4) Software-based \& Non-privileged} &
        \textbf{Description:} Software-only implementations without elevated system-wide privileges. \newline
        \textbf{Characteristics:} (i) Often rely on virtualization layers or application-level sandboxing; (ii) isolate specific processes/data without full system control. &
        \textbf{Open/Academic:} Overshadow; Virtual Ghost; InkTag; Flicker; TrustVisor; Multizone; Utango. \\
        \bottomrule
    \end{tabularx}
\end{table*}

\noindent \textbf{Challenge II: Resource-Aware Engineering.} Training and deploying FMs for autonomous driving tasks demand substantial computational resources, often necessitating large clusters of GPUs or TPUs~\cite{hoffmann2022an, ran2025foundation}. This challenge is significantly exacerbated during in-vehicle deployment, where models must operate efficiently across diverse and resource-constrained hardware platforms. This is a fundamental infrastructure challenge for researchers and developers, opening up the following opportunities:
\begin{itemize}[leftmargin=15pt]
    \item \textbf{Opportunity: Distributed and Collaborative Training.} One key opportunity lies in the development of distributed and collaborative training frameworks~\cite{lin2018deep, moritz2018ray}. These frameworks could enable multiple smaller computing entities to pool their computational resources, meeting the demands of training large-scale FMs. This approach aims to overcome the barrier of high-cost infrastructure but also promote a more diverse and inclusive development ecosystem~\cite{ran2025foundation}.

    \item \textbf{Opportunity: Resource-Aware Model Search for Foundation Models.} Given the diverse and resource-constrained hardware present in autonomous vehicles, resource-aware model search might be a potential direction for efficient deployment. This approach, often leveraging techniques such as neural architecture search (NAS)~\cite{elsken2019neural}, aims to automatically discover specialized model architectures that optimally balance performance with certain constraints like latency, memory footprint~\cite{gao2025ranas}. 
\end{itemize}

\subsection{Development and Deployment of FM4AD}\label{sec:dev_deploy}
Due to their formidable size and computational demands, developing and deploying FMs has posed significant new challenges for autonomous driving applications. In this section, we discuss the challenges and opportunities related to the development and deployment of FMs for autonomous driving.

\noindent \textbf{Challenge I: Understanding the FM Development Toolchain.} The development toolchain for Foundation Models presents unprecedented complexity compared to traditional deep learning frameworks. The enormous scale of these models significantly amplifies the intricacy of data pre-processing pipelines, distributed training systems, and model deployment workflows~\cite{wang2024large}. Compounding this issue is the rapid pace of innovation in the field, which hinders developers and researchers from maintaining a comprehensive understanding of the continuously evolving ecosystem of tools and libraries~\cite{wang2024large}. This complexity and opacity create significant opportunities to systematically analyze and improve how FMs are built and maintained.

\begin{itemize}[leftmargin=15pt]
    \item \textbf{Opportunity: Empirical Analysis for Toolchain Optimization.} A key research opportunity lies in the large-scale empirical analysis of the FM development toolchain. By systematically examining public repositories, development workflows, and the usage of popular libraries, researchers can identify common inefficiencies, performance bottlenecks, and resource-intensive anti-patterns~\cite{wang2024large, nguyen2019machine, liu2023rise}. The insights gained from such studies can lead to data-driven best practices and automated tools that help developers streamline complex processes, optimize resource consumption, and accelerate the overall development lifecycle.
    \item \textbf{Opportunity: Quality Assurance for the FM Development Toolchain.} While traditional deep learning frameworks benefit from a mature suite of quality assurance and testing techniques~\cite{xiao2022metamorphic,wei2022free,zhang2025deep}, these methods are often inadequate for the complex, distributed nature of the FM toolchain. Core components for large-scale training, such as Ray, introduce immense challenges related to network failures, asynchronous operations, and state management that conventional debuggers and testing methods cannot easily handle. Although initial research has begun to address these challenges~\cite{lu2025trainverify, wang2025d3}, these efforts are still in their early stages, leaving a critical opportunity to develop a new generation of QA tools and methodologies specifically for FM engineering. Research in this area could focus on scalable debuggers for distributed systems, fault injection frameworks to test resilience, and novel validation techniques for massive data-parallel pipelines, ultimately enhancing the reliability and reproducibility of the entire FM development process.
\end{itemize}

\noindent \textbf{Challenge II: Efficient and Reliable FM Deployment in Vehicle.} With growing concerns over data privacy and the strict low-latency requirements of autonomous driving, the on-board deployment of FMs in vehicles is becoming essential. A significant body of research has focused on making this feasible through model compression techniques like pruning~\cite{ma2023llmpruner,sun2024a}, knowledge distillation~\cite{gu2024minillm,kim2024promptkd}, and quantization~\cite{lin2024awq,xi2025coat,li2025svdquant}, as well as inference optimizations such as parallel computation~\cite{shoeybi2020megatronlm,yu2024twinpilots} and KV cache management~\cite{kwon2023efficient,luohe2024keep}. However, while these techniques improve efficiency, they can inadvertently introduce new security and reliability vulnerabilities. For instance, researchers have demonstrated tailored attacks that specifically exploit the characteristics of quantized models~\cite{zhang2025verification,yang2024quantization} and KV cache optimizations~\cite{song2025early, wu2025prompt}. This inherent tension between performance and security creates an urgent need for deployment strategies that are both highly efficient and fundamentally trustworthy.
Furthermore, to navigate this tension and help practitioners better trade-off in practice, we provide the risk-utility guidelines (shown in~\Cref{tab:risk_assessment}).

\begin{table*}[!htbp]
\centering
\setlength{\tabcolsep}{4pt} 
\caption{Risk-Utility assessment framework for FM deployment optimization techniques}
\label{tab:risk_assessment}
\begin{tabular}{cp{3.5cm}p{4cm}p{4cm}}
\toprule
\textbf{Technique} & \textbf{Benefit} &\textbf{Introduced Risks} & \textbf{Mitigations}\\
\midrule
Quantization & Maximize speed and storage reduction; hardware compatibility;  no re-training (generally); & Increases susceptibility to attacks targeting quantized models~\cite{zhang2025verification, yang2024quantization, egashira2024exploiting}; increases the chance of hallucination~\cite{li2024dawn} & Safety patching~\cite{chen2025assessing}; safety-aware quantization-aware training; \\
\cmidrule{2-4}

Pruning     & Reduction in computation and memory usage; less risk of hallucination~\cite{chrysostomou2024investigating} & Pruning-activated attacks~\cite{kazuki2026fewer}; privacy leakage~\cite{shang2026msg, kuang2025unveiling} & Security-aware calibration; model patching with repaired parameters~\cite{kazuki2026fewer}; iterative compensation~\cite{frantar2023sparsegpt} \\
\cmidrule{2-4}
Model distillation &  Flexible student model architecture; high performance retention;   & Exploitation of imperfections; teacher hacking~\cite{tiapkin2025on} (transfer of unsafe behaviors, potential lack of generalization);   & Prioritize online data generation (dynamic sampling) and high prompt diversity; limiting re-training epochs; \\
\cmidrule{2-4}

KV Cache Optimization & Enhances real-time responsiveness and throughput; & Potential verbatim input reconstruction and semantic exfiltration through side channel attack~\cite{wu2025prompt}; direct inversion/collision attack~\cite{luo2025shadow} & Obfuscation schemes like KV-Cloak~\cite{luo2025shadow}; execution isolation within TEEs to prevent leakage; \\

\bottomrule
\end{tabular}
\end{table*}

\begin{itemize}[leftmargin=15pt]
    \item \textbf{Opportunity: Empirical Analysis on the Impact of Optimization Techniques.} While specific vulnerabilities in optimized models have been identified~\cite{zhang2025verification,yang2024quantization,song2025early, wu2025prompt}, there is currently a lack of large-scale empirical studies that systematically measure how these techniques impact a model's overall trustworthiness in autonomous driving tasks. This creates a vital research opportunity to conduct comprehensive analyses that quantify the effects of pruning, quantization, and other optimizations on key properties beyond performance. Such studies should evaluate the trade-offs concerning adversarial robustness, fairness, reliability on out-of-distribution data, and the model's propensity for hallucination. The findings would establish an evidence-based understanding for practitioners, leading to clear guidelines for safely applying optimization techniques in safety-critical systems like autonomous vehicles.    
    \item \textbf{Opportunity: Robust-by-Design Optimization Techniques.} A key research opportunity lies in co-designing model optimization techniques with security and robustness as first-class objectives. Instead of optimizing for performance alone, future research could focus on developing new pruning, distillation, or quantization algorithms that are inherently resistant to known attack vectors. For example, this could involve creating quantization schemes that provably maintain adversarial robustness or knowledge distillation processes that transfer security properties from a larger teacher model to a smaller student model. The goal is to create a new class of optimization methods where efficiency gains do not come at the cost of safety.
\end{itemize}

\noindent \textbf{Challenge III: Model Maintenance and Lifecycle Management.} The rapid evolution and high computational cost of FMs create a critical need for structured maintenance and lifecycle management within the ADS development process. A major engineering hurdle is the resource-intensive nature of model training; it is often computationally prohibitive to retrain large-scale models from scratch to adapt to new environmental shifts or specific task variations. Current practices frequently lack systematic frameworks for model versioning, discovery, and reuse. This results in significant redundant effort and inefficiency within the FM supply chain~\cite{wang2024large}, as developers may lack the tools to identify and retrieve existing pre-trained models with similar capabilities that could be adapted through modular updates or lightweight fine-tuning rather than full retraining~\cite{ren2023deeparc}.

\begin{itemize}[leftmargin=15pt]
    \item \textbf{Opportunity: Model Recommendation and Reuse Frameworks.} To mitigate the maintenance overhead, there is a vital research opportunity in developing automated model recommendation engines for the FM-based ADS ecosystem. These frameworks could utilize metadata and gradient-based fingerprinting techniques, such as TENSORGUARD~\cite{wu2025gradient}, to help developers discover similar FMs within a curated repository that best match a target domain's requirements. By treating FMs as software artifacts requiring provenance tracking, researchers can enable similarity detection and family classification independently of training data or specific model formats~\cite{wu2025gradient}.
\end{itemize}

\noindent \textbf{Challenge IV: Edge/Cloud Collaboration for FM Services.} Deploying FMs in autonomous vehicles presents a fundamental trade-off between on-board (edge) and remote (cloud) computations. While optimization techniques like quantization can reduce the burden on edge devices, their limited computing power inherently caps model capability and complexity. Conversely, cloud infrastructure offers vast computational resources for complex reasoning but cannot meet the strict low-latency, high-reliability, and data privacy requirements essential for safety-critical driving decisions. This gap, where neither edge nor cloud alone provides a complete solution, creates a significant opportunity to design hybrid systems that strategically leverage the strengths of both environments.

\begin{itemize}[leftmargin=15pt]
    \item \textbf{Opportunity: Intelligent Task Orchestration.} A primary opportunity lies in creating intelligent frameworks that dynamically schedule and offload tasks across the edge-cloud continuum~\cite{yang2024human, hao2024hybrid}. Research in this area focuses on developing algorithms that can partition workloads in real-time: latency-critical functions like immediate hazard detection would remain on the edge, while computationally intensive, non-real-time tasks like complex scene interpretation or HD map updates could be sent to the cloud. The goal is to build an adaptive system that optimizes for performance, latency, and resource utilization based on current driving context and network conditions.
    \item \textbf{Opportunity: Robust and Asynchronous Communication.} The connection between a vehicle and the cloud is often intermittent and variable. A key opportunity is to design robust and asynchronous communication schemes that ensure system reliability despite unstable network connectivity~\cite{zhang2024assessing}. This includes developing mechanisms that allow the edge model to operate autonomously when disconnected and then asynchronously sync its knowledge or receive updates from the cloud when a connection is re-established. Such schemes are vital for ensuring the vehicle remains safe and operational at all times while still benefiting from the power of the cloud.

\end{itemize}

\section{Foundation Models in Vehicle}
\label{sec:fm4ad}
In this section, we examine how FMs enhance different modules of autonomous driving, summarizing techniques and methodological advances. Specifically, we mainly focus on how FMs can help achieve human-like driving using LLMs, VLMs, and world model-based prediction. We also identify key challenges and research opportunities to guide future investigations in this rapidly evolving field.

\noindent \textbf{Challenge I: Hallucination.} While FMs (i.e., LLMs and VLMs) have achieved significant advancements in autonomous driving, hallucination remains a critical challenge for their safe real-world deployment. Hallucination refers to the generation of outputs that are factually incorrect, inconsistent, or nonsensical, a phenomenon to which FMs are particularly prone~\cite{sun2025hallucination, chakraborty2025hallucination}. Following the formal taxonomy established in recent research~\cite{chakraborty2025hallucination}, these failures can be categorized into four core characteristics: (1) \textbf{Compliance}, where the generation violates non-negotiable hard constraints (e.g., executing an illegal maneuver); (2) \textbf{Desirability}, where the output fails to meet soft constraints measured by optimization objectives like fuel efficiency or passenger comfort; (3) \textbf{Relevancy}, where the model introduces extraneous or off-topic details that do not belong to the driving task; and (4) \textbf{Plausibility}, which assesses the syntactic soundness of the output. Crucially, a hallucination may appear highly plausible and believable to a human critic while still being non-compliant or undesirable, making it a severe risk for safe deployment. \Cref{tab:hallucination_metrics} summarizes the representative hallucination metrics across various task settings, adapted from \cite{chakraborty2025hallucination}. In the context of autonomous driving, a hallucinated object detection, such as mistakenly identifying a non-existent pedestrian, could trigger a severe safety incident like an abrupt stop or a potential collision. Although substantial research has addressed hallucination in general-purpose LLMs~\cite{song2024luna, yao2024llmlieshallucinationsbugs, huang2025survey, huang2024opera, zhou2024analyzing, arteaga2024hallucination} and in FMs for autonomous driving~\cite{dona2024llms,fan2024hallucinationeliminationsemanticenhancement}, the underlying triggers and effective detection methods remain largely unclear, creating vital opportunities for further research.

\begin{itemize}[leftmargin=15pt]
    \item \textbf{Opportunity: Hallucination Detection and Mitigation.} While the underlying mechanism of hallucination is currently unclear, the first opportunity lies in the detection and mitigation of hallucination. To combat hallucination, research efforts have been made by both the AI and SE communities. For instance, Yang~\textit{et al.}~\cite{yang2025hallucination} utilise metamorphic relations to detect hallucinations in LLMs. Future research could focus on developing detection and mitigation strategies that reduce hallucinations without hurting the model's performance.
    \item \textbf{Opportunity: Multi-modal Grounding and Verification.} One promising direction to tackle hallucination is to leverage the inherently multi-modal nature of the perception data. Research may focus on developing methods to ground an FM's generated output (e.g., textual caption) against raw sensor information from cameras, LiDAR, or radar~\cite{chakraborty2025hallucination}. Developing robust grounding techniques would anchor the model's outputs to physical reality, which can significantly reduce the likelihood of factually incorrect or unverified statements. 
\end{itemize}

\begin{table*}[!htbp]
\centering
\footnotesize
\caption{Summary of Hallucination Metrics across Task Settings (Adapted from~\cite{chakraborty2025hallucination})}
\label{tab:hallucination_metrics}
\begin{tabularx}{\textwidth}{@{}lXXXX@{}}
\toprule
\textbf{Task Setting} & \textbf{Compliance Metrics} & \textbf{Desirability Metrics} & \textbf{Relevancy Metrics} & \textbf{Plausibility Metrics} \\ \midrule
\textbf{Question-Answering} & Accuracy, FactScore, BERTScore, Contradictions & Calibration Error, Succinctness, Prudence & Cross Encoder, Perplexity & ROUGE, Semantic Preciseness \\ \midrule
\textbf{Image Captioning} & CHAIR, POPE, CLIP Score, METEOR & CIDER Human Alignment, Detailedness & CHAIR, POPE & ROUGE, SPICE, Perplexity \\ \midrule
\textbf{Planning} & Feasibility, Action Probabilities, Plan/Action Accuracy & Success Rate, Clarification Rate, Unsafe Action Rate, Calibration Error, Estimated Payoff & Action Probabilities & Non-compliance Rate, Preciseness \\ \bottomrule
\end{tabularx}
\end{table*}

\noindent \textbf{Challenge II: Multi-modality Adaptation.} While Foundation Models (FMs), particularly Large Language and Multi-modal Large Language Models, have demonstrated remarkable reasoning capabilities in autonomous driving, their practical application is often hindered by a critical vulnerability. Most current approaches rely heavily on the processed outputs of upstream perception modules, treating them as absolute ground truth~\cite{gao2025survey}. This tight coupling means the entire system is brittle; even minor perception inaccuracies, such as a slight error in an object's heading estimation, can cascade and lead to catastrophic failures in the downstream decision-making process~\cite{mao2024a}. This dependency on imperfect perception highlights a critical need for robust adaptation methods, opening up several research opportunities.

\begin{itemize}[leftmargin=15pt]
    \item \textbf{Opportunity: Benchmarking Robustness in Multi-Modal Fusion.} As research moves towards end-to-end models, there is a pressing need for standardized benchmarks to rigorously evaluate the quality and robustness of their sensor fusion capabilities. Current evaluations often focus on overall task performance, making it difficult to isolate how well a model handles sensor noise, failure, or conflicting information. This creates a significant opportunity to develop novel evaluation protocols and challenging datasets designed specifically to probe the fusion process. The creation of these benchmarks is critical for systematically comparing different architectures and guiding the development of more reliable and truly robust multimodal systems.

    \item \textbf{Opportunity: End-to-End Multi-Modal Fusion.} A promising research direction is to move beyond cascaded pipelines and develop integrated, end-to-end FMs. The goal is to create architectures that can directly ingest and fuse raw, heterogeneous sensor data (e.g., camera pixels, LiDAR point clouds, radar signals). By learning directly from raw data, the model can develop its own robust internal representations, bypassing the vulnerabilities of a brittle intermediate perception layer. This approach allows the FM to learn complex cross-modal correlations, enabling it to rely on one modality to compensate for noise or errors in another.

\end{itemize}

\noindent \textbf{Challenge \& Opportunity: Domain-Specific Foundation Models for Autonomous Driving.} While the open-source landscape for code-centric large language models (LLMs) has thrived with examples like Magicoder~\cite{wei2024magicoder} and CodeLlama~\cite{reziere2024codellama} setting benchmarks, most FMs for autonomous driving remain proprietary, such as GAIA-1~\cite{hu2023gaia1generativeworldmodel}. This restricts academic and independent researchers from advancing innovation in a field where safety and robustness are critical. The core problem is the absence of a pre-trained, powerful foundation model designed for the specific tasks of autonomous driving, such as integrating multi-modal data (e.g., cameras, LiDAR) and handling complex decision-making in dynamic environments. An open-source, domain-specific foundation model is urgently needed to bridge this gap. By providing a robust starting point for tasks like perception, planning, and control, this model would empower researchers to address real-world driving challenges efficiently.

\subsection{FM-Enabled Intelligent User Experience}\label{sec:intelligent_user}
Beyond enhancing core autonomous driving capabilities, FMs are revolutionizing user interaction and experience in autonomous vehicles. This subsection explores two key aspects: intelligent user interfaces with personalization, and enhanced surrounding awareness capabilities.

\noindent \textbf{Challenge \& Opportunity: Intelligent User Interface and Personalization.} While FMs enable more intelligent and personalized user experiences in autonomous vehicles, several challenges need to be addressed. MLLMs like GPT-4V can interpret natural language instructions to control vehicles according to user preferences. For example, Cui \emph{et al.} demonstrated that LLM-based planners can respond to personalized commands such as ``drive aggressively,'' adjusting vehicle behavior across different speeds and risk levels~\cite{cui2024receive}. However, this flexibility raises significant safety concerns. As shown in~\cite{cui2024personalized}, LLMs may interpret and execute potentially dangerous commands like ``drive as fast as you can.'' Although research has explored methods to ensure compliance with traffic rules and safety requirements~\cite{cui2024personalized, yang2024human}, the vulnerability to jailbreak attacks remains a concern, particularly given the proliferation of LLM exploitation techniques~\cite{rao2024tricking,zou2023universal,zhu2024autodan, jin2024jailbreakzoosurveylandscapeshorizons}. Additionally, balancing real-time responsiveness with user privacy presents another significant challenge, as discussed in Section~\ref{sec:dev_deploy}.

\noindent \textbf{Challenge \& Opportunity: FM-Enabled Surrounding Awareness.} FMs could enhance surrounding awareness by providing users with real-time, interpretable insights about the vehicle's environment. For instance, DriveGPT4~\cite{xu2024DriveGPT4} integrates this awareness into the driving loop, offering passengers explanations for vehicle actions (e.g., ``veers left to avoid collision''). This awareness extends to both safety and convenience features, such as alerting users to nearby hazards or points of interest, enhancing the overall experience~\cite{domova2024comfort, miller2014situation}. However, ensuring the accuracy and reliability of FM-generated insights remains challenging, as hallucinations or misinterpretations could mislead users. Additionally, presenting complex information requires careful UI design to maintain user-friendliness. Potential opportunities include developing robust multi-modal grounding techniques to reduce errors through cross-validation of visual and textual data, and creating intuitive visualization methods such as augmented reality overlays to effectively convey FM insights. These advancements could transform vehicles into intelligent companions that enhance both safety and user engagement.

\section{Foundation Model Application in Practice}
\label{sec:practice}

This section explores the practical deployment of FMs in autonomous driving. We distinguish between modular integration and full adoption of FMs, showcasing their role in enhancing vehicle capabilities. Besides, we also articulate their application in automating the development of autonomous driving systems, along with the potential challenges and opportunities.

Several initiatives employ FMs as specialized components within autonomous driving systems. Xiaomi SU7 integrates a VLM via OTA update to enhance scene interpretation and safety alerts~\cite{su7}. Li Auto combines a VLM with an end-to-end framework in its OTA-updated smart driving system, improving scene recognition and maneuver accuracy~\cite{liauto}. TIER IV utilizes an LLM to enable vehicles to reason and communicate, enhancing human-vehicle interaction~\cite{tier4}. Similarly, Bosch researchers apply natural language processing to predict traffic behaviors, boosting situational awareness~\cite{keysan2023can}. These cases demonstrate FMs augmenting specific functions like perception and communication.

Meanwhile, full adoption leverages FMs as the core of autonomous driving systems. Cui et al. deploy an LLM in Talk2Drive for end-to-end control, personalizing driving through language and vision inputs \cite{cui2024personalized}. Their subsequent work fully integrates a VLM onboard for motion control, unifying perception and decision-making \cite{cui2024onboardvisionlanguagemodelspersonalized}. Waymo’s MotionLM uses FMs to transform multi-agent motion prediction into a language task, streamlining dynamic interactions \cite{seff2023motionlm}. These efforts highlight FMs driving comprehensive, adaptive autonomy.

\noindent \textbf{Challenge I: Foundation Model Alignment.} As FMs become increasingly integrated into autonomous driving systems, their potential societal risks demand careful consideration. The undesired behaviors exhibited by FMs, such as hallucination, raise particular concerns in safety-critical domains like autonomous driving where they directly impact public safety. AI alignment has emerged as a potential solution, aiming to ensure AI systems behave in accordance with human intentions and values~\cite{leike2018scalableagentalignmentreward}. Despite its critical importance for the safe deployment of FMs in autonomous driving, research in this area remains limited~\cite{kong2024superalignment, ieee2024standard, abbo2024social, wolf2024fundamental}. The complexity of foundation model systems, encompassing fairness, privacy, and security concerns, urgently calls for more attention and investigation into alignment strategies. Current alignment research can basically be divided into two key components: forward alignment and backward alignment~\cite{ji2025aialignmentcomprehensivesurvey}. Below are potential opportunities:

\begin{itemize}[leftmargin=15pt]
    \item \textbf{Opportunity: Enhancing Feedback Mechanisms (Forward Alignment).} Forward alignment, which focuses on proactively shaping model behavior during training, presents a significant opportunity for improving FMs in autonomous driving. By incorporating human-value feedback during the training process, developers can construct more robust systems where FMs not only continuously learn but also maintain alignment with human intentions and safety requirements~\cite{wang2024large}.

    \item \textbf{Opportunity: Safety Benchmarks and Evaluation for Assurance (Backward Alignment).} Datasets and benchmarks are crucial for safety evaluation, serving as fundamental tools for ensuring AI alignment. A key opportunity lies in developing comprehensive metrics and benchmarks for FMs to better evaluate their safety performance and ability to minimize accidents during task execution~\cite{ji2025aialignmentcomprehensivesurvey}. Unlike traditional deep learning models, FMs can leverage general knowledge rather than actual cues to achieve unexpectedly high scores on existing metrics~\cite{xie2025are}. This limitation highlights the need for more comprehensive benchmarks and metrics that can accurately assess both FM capabilities and potential deviations from intended behaviors~\cite{amodei2016concreteproblemsaisafety}.

    \item \textbf{Opportunity: Online Methods for Safety Alignment.} While safety benchmarks can partially ensure alignment, online methods are still needed to guarantee safety during task execution in autonomous vehicles. However, due to the complexity of FMs and the requirement for low-latency execution, designing such methods remains challenging. As highlighted by OpenAI, alignment must embrace uncertainty while maintaining rigorous measurement~\cite{openai4alignment}. This limitation highlights the need for designing suitable online methods for FMs in autonomous driving systems.
\end{itemize}

\noindent \textbf{Challenge II: Ethical and Human-Like Autonomous Driving.} The ultimate objective of integrating FMs into autonomous driving is to enable ethical and human-like driving behavior. However, as pointed out by Tian~\emph{et al.}~\cite{tian2026large}, existing regulatory frameworks for autonomous vehicles (e.g., ISO 26262) primarily address functional safety and cybersecurity for conventional rule-based systems, and do not fully capture the complexities introduced by foundation models; thus, these regulations may need to be revisited and updated. Moreover, transparency and interpretability in ADS have been increasingly emphasised due to their critical roles in regulatory approval and accountability. The integration of FMs may offer new opportunities for interpretability---for example, in speech-to-speech systems, researchers can inspect intermediate textual representations produced by LLMs to facilitate error analysis~\cite{pan2026s2st}. Nevertheless, further advances are still required to systematically embed explainability into FM-driven ADS pipelines. Finally, since FMs are often adopted for reasoning and decision-making, it is crucial to test them under morally sensitive scenarios and to characterise how risks are distributed among different road users. Recent work has begun to operationalise such moral testing via simulation-based methods (e.g., metamorphic testing) to expose potentially problematic decision patterns~\cite{tang2025moral}. These efforts are essential for developing ethical and human-like autonomous driving.

\begin{itemize}[leftmargin=15pt]
    \item \textbf{Opportunity: Moral Testing and Assurance for FM-based ADS.} As FMs are increasingly adopted for high-level reasoning and decision-making in ADS, a key opportunity is to establish systematic \emph{moral testing} and assurance pipelines that complement functional safety validation. Unlike conventional safety requirements, moral expectations are often value-laden, context-dependent, and may not admit a single ground-truth label, making classic test oracles difficult to define. Recent work has started to operationalise moral evaluation via simulation-based \emph{metamorphic testing}, where moral meta-principles are encoded as relational properties across paired scenarios to expose inconsistent or discriminatory decision patterns~\cite{tang2025moral}. Building on this direction, future research could develop scalable moral test suites and benchmarks, integrate scenario generation and coverage metrics tailored to morally sensitive situations.
\end{itemize}

\noindent \textbf{Challenge III: Ensuring Correctness of FM-Generated Code.} With the widespread application of FM-based coding assistance (i.e., GitHub Co-pilot), autonomous driving developers will inevitably use FM-generated code during the development of the autonomous driving system. As pointed out by Chen~\textit{et al.}~\cite{chen2025deepdivelargelanguage}, LLM could misunderstand the code description and generate code without syntactic mistakes, but still defective. Furthermore, code generated by LLMs often lacks description or context and is more difficult for human developers to understand and maintain~\cite{liu2024refining}, and may introduce critical issues into the codebase if not carefully verified~\cite{molison2025llmgeneratedcodemaintainable}.

\begin{itemize}[leftmargin=15pt]
    \item \textbf{Opportunity: Automated Verification for FM-generated Code.} With the growing coding capabilities of LLMs, the use of LM-generated code is becoming inevitable. However, errors in such code can be more difficult for human developers to detect. Therefore, in safety-critical domains like autonomous driving, there is a pressing need to develop automated verification techniques for FM-generated code. Nouri~\textit{et al.}~\cite{nouri2025simulation} proposed an iterative loop to automatically generate, verify, and refine LLM-generated code in the context of autonomous driving. Yet, due to issues such as LLMs’ sensitivity to minor text disruptions in code descriptions or prompts, more robust automated verification techniques are still required.
\end{itemize}

\section{Future Research Roadmap}\label{sec:roadmap}
The development and deployment of Foundation Models for Autonomous Driving (FM4AD) is a monumental task that requires a coordinated, multi-stage research effort. To guide the community, we propose a roadmap divided into three phases: short-term foundational work, mid-term system integration, and a long-term vision for achieving verifiable trustworthiness at scale.

\subsection{Short-Term Goal}
The immediate focus must be on establishing the fundamental building blocks required for robust and scalable research and development of FMs for autonomous driving. This involves addressing the most pressing challenges in data, development tools, and deployment efficiency.

\begin{itemize}[leftmargin=15pt]
    \item \textbf{Establishing Data Best Practices:} The quality and management of data are paramount. The community's short-term goals should be to develop \textbf{unified data management frameworks} to handle the massive scale of autonomous driving datasets and to create automated tools for \textbf{dataset license compliance} to ensure legal and ethical usage. Simultaneously, advancing techniques for \textbf{bias mitigation} and \textbf{privacy preservation} during data cleaning and curation is critical to building a fair and secure foundation. Success in this phase will be measured by the development of automated fairness testing suites that achieve a high correlation between perception-level bias metrics (e.g., AOD) and decision-level safety violations.

    \item \textbf{Improving the Development Toolchain:} To manage the unprecedented complexity of the FM toolchain, the immediate priority is to conduct \textbf{large-scale empirical analyses} to identify and eliminate inefficiencies in current workflows. A parallel effort is required to develop a new generation of \textbf{Quality Assurance} and testing techniques specifically designed for the distributed systems (e.g., Ray) that underpin FM training, thereby improving the reliability and reproducibility of the development process.

    \item \textbf{Enabling Efficient and Safe Deployment:} Before FMs can be widely integrated, their efficiency and safety on resource-constrained vehicle hardware must be understood and improved. A crucial short-term task is to perform a comprehensive \textbf{empirical analysis of the impact of optimization techniques} like quantization and pruning on model trustworthiness. This knowledge will directly inform the development of \textbf{robust-by-design optimization} methods, where security and reliability are co-designed with performance. Furthermore, success in this area will be measured by the ability of optimized models to maintain an inference latency within the 500ms range, a critical threshold for ensuring real-time responsiveness in complex driving reasoning tasks~\cite{tian2026large}.
\end{itemize}

\subsection{Mid-term Goal}
With a solid foundation in place, the research focus can shift to the complex challenges of integrating FMs into the vehicle as a cohesive and reliable system. This phase emphasizes system-level reliability and architecture.

\begin{itemize}[leftmargin=15pt]
    \item \textbf{Enhancing System Reliability and Adaptation:} The mid-term priority is to address the core reliability issues of FMs in dynamic environments, namely \textbf{hallucination} and \textbf{multi-modality adaptation}. This involves developing \textbf{end-to-end multi-modal fusion} models that are less vulnerable to upstream perception errors and using \textbf{multi-modal grounding} to verify model outputs against raw sensor data. To measure progress, creating \textbf{standardized benchmarks for robustness in multi-modal fusion} is essential.

    \item \textbf{Designing Hybrid System Architectures:} An integrated FM-powered vehicle will not operate in isolation. A key mid-term goal is to architect effective \textbf{edge-cloud collaboration} systems. Research must deliver frameworks for \textbf{intelligent task orchestration} and \textbf{robust, asynchronous communication} schemes that can dynamically balance the low-latency needs of the vehicle (edge) with the immense computational power of the cloud.

    \item \textbf{Fostering a Domain-Specific Ecosystem:} To accelerate innovation and democratize research, the community must address the lack of powerful, open-source models. A central mid-term objective should be the development of an \textbf{open-source, domain-specific foundation model for autonomous driving} that can serve as a strong baseline for academic and independent researchers.
\end{itemize}

\subsection{Long-term Goal}
The ultimate vision is the widespread deployment of autonomous vehicles powered by FMs that are not just capable but are verifiably safe, trustworthy, and aligned with human values. This requires tackling the most fundamental challenges from the hardware to the AI's core objectives.

\begin{itemize}[leftmargin=15pt]
    \item \textbf{Establishing a Hardware Root of Trust:} Trustworthiness must begin at the silicon level. The long-term grand challenge is to overcome the risks of "black-box" hardware by developing \textbf{security-by-design hardware architectures}. This includes creating \textbf{Trusted Execution Environments (TEEs)} tailored for AI accelerators and promoting \textbf{open and verifiable hardware stacks} that allow for community-driven security audits, providing a transparent and secure foundation for all software.

    \item \textbf{Achieving Foundation Model Alignment:} Ensuring that FMs behave in accordance with human intentions is perhaps the most critical long-term goal. This demands a sustained research program in \textbf{AI alignment}. The focus should be on developing robust methods for both \textbf{forward alignment}, such as enhancing human-feedback mechanisms during training, and \textbf{backward alignment}, by creating comprehensive \textbf{safety benchmarks} to evaluate model behavior. A key frontier is the design of effective \textbf{online methods for safety alignment} that can provide real-time assurance in a deployed vehicle.

    \item \textbf{Guaranteeing Code and System Correctness:} As FMs increasingly contribute to writing the software that runs autonomous vehicles, verifying its correctness is a long-term imperative. The research community must work towards powerful \textbf{automated verification techniques} that can formally prove the safety and reliability of \textbf{FM-generated code}, ensuring that this powerful tool enhances, rather than compromises, the safety of autonomous systems.
\end{itemize}

\section{Conclusion}
\label{sec:conclusion}

In this paper, we presented a structured literature review and an initial roadmap for integrating foundation models into autonomous driving systems. We organized the discussion around three dimensions: FM infrastructure, in-vehicle integration, and practical deployment. Across these dimensions, we synthesized the state of the art, identified key challenges, and highlighted promising research opportunities, which we further distilled into short-, mid-, and long-term research goals. Although substantial technical and engineering challenges remain, FMs offer significant potential to advance the capabilities of ADSs. We hope this paper serves as a useful reference for future research and helps guide the development of safer, more reliable, and more trustworthy autonomous driving systems.


\begin{acks}
This work was partially supported by the National Natural Science Foundation of China: No. 52408039, JST CRONOS Grant (No. JPMJCS24K8) and JSPS KAKENHI Grant (No.JP24K02920). This research/project was also supported by the Ministry of Education, Singapore under its Academic Research Fund Tier 2 (Proposal ID: T2EP20223-0043; Project ID: MOE-000613-00). Any opinions, findings and conclusions or recommendations expressed in this material are those of the author(s) and do not reflect the views of the Ministry of Education, Singapore.
\end{acks}

\bibliographystyle{ACM-Reference-Format}
\bibliography{main}










\end{document}